# Optical tweezers approaches for probing multiscale protein mechanics and assembly

Kathrin Lehmann,[1,2] Marjan Shayegan,[3] Gerhard A. Blab[2] and Nancy R. Forde[1,4,5,6]

[1]Department of Physics, Simon Fraser University

[2]Soft Condensed Matter & Biophysics, Utrecht University

[3]School of Engineering and Applied Sciences, Harvard University

[4]Department of Molecular Biology and Biochemistry, Simon Fraser University

[5]Department of Chemistry, Simon Fraser University

[6]Centre for Cell Biology, Development and Disease (C2D2), Simon Fraser University

Multi-step assembly of individual protein building blocks is key to the formation of essential higher-order structures inside and outside of cells. Optical tweezers is a technique well suited to investigate the mechanics and dynamics of these structures at a variety of size scales. In this mini-review, we highlight experiments that have used optical tweezers to investigate protein assembly and mechanics, with a focus on the extracellular matrix protein collagen. These examples demonstrate how optical tweezers can be used to study mechanics across length scales, ranging from the single-molecule level to fibrils to protein networks. We discuss challenges in experimental design and interpretation, opportunities for integration with other experimental modalities, and applications of optical tweezers to current questions in protein mechanics and assembly.

**Introduction**

Biology has evolved proteins capable of self-assembly that create dynamic scaffolds imparting mechanical stability and force responsiveness inside and outside of cells. Intracellular proteins include actin, tubulin, tropomyosin and titin, which contribute to cytoskeletal and muscle structure and mechanics. Extracellularly, proteins including collagen, elastin and fibrin assemble to form the extracellular matrix and connective tissues. Understanding how these higher-order assemblies of proteins achieve their responsive mechanical functions requires the ability to measure their mechanical response in different chemical environments and at different hierarchical levels of organization. Because the properties of these proteins are encoded at the molecular level, and because mechanics of the higher-order assemblies can be drastically altered by molecular changes in composition (e.g. mutations, post-translational modifications, age-related chemical changes or ligand binding) [1-6], it is important to characterize mechanical response starting from the single-molecule level.

Various techniques have been developed to perform single-molecule force spectroscopy (SMFS). These include atomic force microscopy (AFM), magnetic tweezers (MT), centrifuge force microscopy (CFM), acoustic force spectroscopy (AFS) and optical tweezers (OT). As described in previous reviews [7-9], these techniques confer distinct advantages, meaning the choice of SMFS technique can be dictated by desire for high spatial, temporal and/or force resolution; force range; high throughput measurements; or application of torque in addition to linear stretching forces. However, only some of these approaches are amenable to characterizing mechanics of higher-order protein structures such as filamentous/fibrillar assemblies and larger-scale networks.

OT have the broadest applicability in probing the mechanics of protein assemblies at various hierarchies of scale: they offer advantages of high spatial, temporal and force resolution for SFMS, and the ability to passively and actively probe microscale mechanics at prescribed locations within three-dimensional protein networks and even inside living cells [10, 11]. This ability to extract force and displacement information across a wide span of system size scales with a single experimental approach facilitates meaningful comparisons of mechanics of proteins at different levels of assembly.

In this mini-review, we highlight some of the applications of OT to the study of protein mechanics, ranging from single-molecule investigations of mechanics and unfolding to studies of higher-order fibrils and networks. We provide examples of studies on collagen throughout, and highlight work on other assembling protein systems including actin filaments, microtubules, fibrin and prions. We close with a brief discussion of prospects for future research.

**Single-molecule investigations of protein mechanics**

SMFS studies of protein mechanics require linking the ends of a protein to larger objects that can be independently manipulated. For OT studies, at least one end is linked to a micron-sized bead, which can be held in the focused laser beam of an optical trap (Figure 1). When stretched by its other end (e.g. via manipulation of a bead held on a movable pipette or in an optical trap, or by moving a glass slide), the displacement of the trapped bead from the focus provides the force applied to stretch, while the separation between the two ends gives the end-to-end extension of the molecule [12]. Thus, the primary read-out from an SMFS study is a force-extension curve (FEC), though other modalities such as constant-force measurements can be used to provide deeper information of folding/unfolding dynamics [12-14]. FECs reveal information about the elasticity and flexibility of a protein, changes in its structure – such as unfolding – induced by force, and the timescales on which structural changes occur.

Proteins can exhibit a variety of responses to applied force. An entropic elastic response arises when randomly coiled or bent structures are straightened by an applied force, without changing internal structure or contour length. Here, deformation is reversible on the timescale of SMFS and energy put into straightening is recovered upon relaxation. The FEC is monotonic and can be described by the worm-like chain (WLC) or freely jointed chain (FJC) model of polymer flexibility (Fig. 1A). When proteins structurally deform, the FEC can exhibit sharp features, seen as "sawtooth" characteristics arising from domain unfolding and accompanying release of previously buried polypeptide backbone to the force-bearing region of the chain (Fig. 1B). The reversibility of these structural changes can be ascertained by looking for hysteresis between stretch and relaxation curves. When pulling occurs more rapidly than internal equilibration of the protein, the loading-rate dependence of unfolding force provides information about the location and heights of free energy barriers, as outlined in [13]. Alternatively to sawtooth signatures of domain unfolding, more gradual changes of contour length may occur, arising from structural distortions along the force-bearing backbone. Such backbone lengthening and structural transitions are exhibited by DNA [15-18], and may also contribute to the force response of collagen, a protein with a 300-nm long triple-helical structure (Fig. 1A). Initial OT-SMFS measurements fit collagen's FEC with the inextensible WLC model with a persistence length of ~15 nm, describing collagen as a relatively flexible polymer [19-22]. More recently, however, measurements of its flexibility using AFM imaging indicate that it is far less flexible, with a persistence length of ~95 nm [23], and low-force structural distortions in collagen have been implicated by a variety of single-molecule approaches, including OT [24], MT [25-27] and CFM [28]. A low-force-induced structural lengthening of collagen's triple helix could reconcile the disparity in persistence lengths: by fitting a FEC over variable force ranges, Rezaei *et al.* found that the WLC persistence length increased significantly as the maximum force used for fitting decreased [24]. Potential mechanisms for a force-induced "softening" of collagen, which may involve bend-twist coupling [29], are described in a recent review [30].

Challenges arise for interpreting molecular flexibility obtained from OT-SMFS measurements when the molecules' contour length is not significantly longer than the persistence length. In this case, the persistence length extracted from WLC fits may significantly underestimate the polymer's true persistence length, as seen for DNA [31, 32]. This underestimation arises from the finite length of the experimental chain: the shorter the chain, the greater the relative importance of contributions from its ends. By tethering the chain by its end to surfaces, the orientations of its ends are restricted, which alters the total conformational entropy of the chain and affects determination of its persistence length[31, 32]. In some cases, the "true" persistence length can be obtained by measuring polymers of different contour lengths and extrapolating results to the infinite-length limit. This approach has been used for measurements on short lengths of DNA [31, 32], whose contour length is easily controlled, but it is not as easily generalizable to proteins such as collagen, whose length is biologically regulated and which may not fold or be secreted properly if lengthened via genetic engineering.

Probing a tethered molecule substantially shorter than the bead diameter also generates experimental challenges. Having two microspheres at separations much less than their sizes (and the size of the trapping laser focus/foci) can lead to optical interference between a bead and the other trap, an effect which must be deconvolved from the response to obtain the desired force readout of tension applied to the molecule [21, 33]. Measurements of such short polymers also suffer from amplified effects of stage drift and off-axis stretching. These shortcomings can be addressed by using DNA "handles" to link the protein ends to beads, thereby extending the separation between particles and avoiding unwanted optical interference (Fig. 1B). DNA handles have become a standard complement of single-molecule measurements of protein folding with optical tweezers [34, 35]. The topic of force-induced protein unfolding, and what can be learned about the energy landscapes, is the subject of many reviews, to which we refer the interested reader [12-14, 36, 37].

OT-SMFS lends itself well to understanding how protein-protein interactions can alter the mechanical landscape. Many proteins retain their structure and independence of folding in the context of neighbouring domains – this feature has been used to recapitulate the force response of titin in muscle from studies of its individual domains [38]. Alternatively, protein interaction can stimulate pathological protein misfolding. OT-SMFS studies have demonstrated that prion proteins coupled together in series do not maintain their independent structure, instead adopting new misfolded structures and unfolding pathways compared with monomeric prions (Fig. 1B) [39-41]. Because OT can be used to unfold and refold the same protein hundreds to thousands of times, rare folding/unfolding events can be captured, events that may be critical for initiating formation of larger-scale misfolded aggregates that lead to disease [42, 43].

**Mechanics of higher-order protein fibres and networks**

An ongoing challenge is to link protein mechanics at the molecular level to the mechanics of higher-order assemblies. What are the energetic hierarchies governing supramolecular response? In the context of a collagen fibril formed from laterally associated molecules (Fig. 1C), for example, what role is played by straightening of collagen's triple-helical backbone (governed by bending rigidity / persistence length at the molecular level) versus molecular deformation (e.g. triple-helix unwinding) versus intermolecular lateral sliding of triple helices? Intermolecular sliding is restricted by covalent crosslinking between collagens, a modification that can be biologically prescribed during assembly and which also can occur, less site-specifically, as tissues age [1, 44-46]. Crosslinking between chains within a triple helix may also alter its ability to deform at the molecular level when stretched (e.g. by locally pinning and therefore preventing unwinding of the three chains) [28, 47], though it is less clear how intramolecular crosslinks may affect the bending persistence length at the molecular level [23, 30]. The challenge of linking mechanics at different length scales becomes greater when bridging to even higher-order bundles and networks (e.g. collagen fibrillar gels; Fig. 2A) [48-53].

Lateral association into higher-order protein fibres produces structures that are stiffer than their individual protein components. The Young's modulus of such structures is significantly larger than the elastic modulus (proportional to stiffness) of typical optical traps, meaning only high-intensity OT have been used to perform meaningful strain measurements on fibres [54-57]. A disadvantage of high-intensity optical traps is the local heating of the sample that can result [7]. With low-intensity OT, bending moduli of the fibres can be determined by pushing an optically trapped bead laterally against the filament and measuring its deformation as a function of applied force (Fig. 1C). OT have been used to measure the bending stiffness of microtubules [58-62], actin filaments [60] and collagen fibrils [63]. Other examples of mechanical properties of protein fibres measured with OT include torsional stiffness [64, 65], bending stiffness of fibre bundles [66, 67], sliding forces of filaments within bundles [68], and bending and spontaneous assembly of two interacting filaments using a four-trap OT instrument [69]. OT can also be used to measure the assembly and disassembly of individual protein filaments, as has been done for microtubules [70, 71].

Larger-scale protein networks can be mechanically probed *in situ* using OT-based microrheology (OT-MR) [72-75]. In OT-MR, the motion of an optically trapped bead (constrained within a trap that is either stationary or actively driven) is used to read out information about the frequency-dependent viscoelasticity of its surroundings (Fig. 2) [72, 75-77]. Particle dynamics can be used to determine mechanical properties of the surrounding network. One such property is the complex shear modulus $G^*(f)$, comprising the storage and loss moduli ($G'(f)$ and $G''(f)$, respectively). These moduli describe the elastic and viscous response of the medium (proteins + solvent) surrounding the particle. The frequency dependence of the moduli provides information on the bending rigidities and interactions between protein filaments in solution [72, 78]. Because of this, care must be taken to correctly determine and correct for contributions from OT trap stiffness, which can be particularly difficult when probing softer networks whose elastic modulus is comparable to that of the optical trap [79]. Additionally, *in-situ* calibration of optical traps is more challenging in these complex media than in the aqueous solutions used for SFMS, due to the local environment being both viscoelastic and locally heterogeneous, and to light scattering from larger-sized components of the network [80-82]. Alternative approaches to MR exist, such as passive observation of bead diffusion, though these generally have a more limited bandwidth and also rely on microscopic tracer beads [72]. The use of beads for these MR approaches creates some requirements in experimental design. Choice of bead size for MR is important: if a continuum measure of network properties is desired, beads should be at least 3x the pore size of the network [75]. Conversely, smaller beads added to a sample are more easily able to navigate through the pores and their motion can reveal information about the sample's spatial heterogeneity. Beads should be sufficiently dilute in the sample so as not to influence the mechanical properties (e.g. forming a much more rigid bead-gel composite structure). It is also essential to characterize and control for nonspecific interactions between the bead surface and the protein network, a concern that should be addressed in all types of bead-based measurements on proteins and protein networks [83, 84]. OT-MR has been widely applied to characterize network mechanics of cytoskeletal proteins (including actin [76, 85-92], intermediate filaments [93, 94], and microtubules [90, 92]) and of extracellular proteins (including collagen [22, 79, 95-97] and fibrin [78, 98-100]).

Most OT-MR studies analyze the motion of a single trapped bead to learn about its local microenvironment, but multiple-particle OT-MR can also be performed. It provides distinct information about the through-space mechanical coupling of the network by analyzing correlated motion between pairs of beads [101]. Often one bead is actively displaced with OT and the motion of other (non-optically trapped) beads in the network is recorded, monitoring for example the amplitude and phase lag of their motion relative to the driven particle, which can be used to determine the mechanical transfer function of the network [81, 94]. For higher-frequency information, several beads can be optically trapped simultaneously, and their correlated motion determined either through active oscillation of one particle or through passive recording of their thermally driven dynamics in stationary traps [77]. Utilizing traps to position beads at desired locations within the network provides greater control over their separations and orientation with respect to the (potentially anisotropic) network [96]. To our knowledge, a maximum of two traps has thus far been employed simultaneously in OT-MR experiments, though more could be implemented with methods such as holographic optical tweezers [96, 102-104].

**Protein network formation and remodelling**

The assembly of proteins from solution into larger-scale networks triggers many changes in the local microenvironment (Fig. 2A), which can be sensed by optically trapped beads. If beads are comparable to the mean pore size in the network, then measurements of local environment may be highly heterogeneous, with some beads sensing essentially solvent while others – more tightly embedded between network filaments – report very high elastic moduli. Studies on collagen assembly into fibrillar networks illustrate the heterogeneous properties sensed by micron-sized beads: following triggering of assembly, the heterogeneity in elastic modulus increases as collagens assemble into fibrils that form networks, eventually reaching a plateau [79, 95]. The kinetics of increase in $G^*(f)$ matches development of turbidity in the sample, indicating commensurate growth in mechanical protein structures and light-scattering fibrillar structures (Fig. 2A) [79]. OT-MR has been applied to study the triggered assembly and disassembly of other protein networks, such as actin and fibrin [91, 100].

During MR measurements, optical tweezers also afford the ability to monitor development of mechanical environment at a given location over time, by repeatedly probing the dynamics of the same particle (Fig. 2A). This particular type of measurement presents technical challenges: forces exerted by the assembling proteins can be sufficiently strong to displace the particle from the trap [79]; and maintaining the optical trap always-on for long periods of time can result in local heating of the sample, which in turn may alter thermally sensitive assembly kinetics and – for proteins with marginal thermal stability such as collagen – even molecular protein structure [7, 105]. Nonetheless, such measurements are possible, and have been used to characterize locally evolving mechanics during network formation [79, 95, 100].

Transient protein-protein interactions, which can provide nucleation points for higher-order assembly, also can be revealed by OT-MR. As one example, collagen assembly into fibrils is facilitated by short non-triple-helical regions at the ends of the collagen molecule, called telopeptides. OT-MR revealed striking enhancement of the viscoelastic properties of solutions of collagens with their telopeptides intact compared with collagens whose telopeptides had been removed (Fig. 2B) [106]. These differences were seen in solution conditions that inhibit lateral assembly into fibrils, suggesting that telopeptides enhance stickiness between collagen chains in a variety of solution conditions, a finding supported by a polymer association model. Furthermore, OT-MR was used kinetically to detect enzymatic alteration of these intermolecular interactions: the viscoelasticity of collagen solutions was found to decrease as enzymes cleaved the telopeptides from the ends of the collagen proteins (Fig. 2B). It may also be possible for OT-MR to determine the kinetics of transient protein-protein interactions contributing to network assembly. As predicted and found in bulk rheology studies of actin networks, a local maximum in the frequency-dependent loss modulus, $G''(f)$, is observed at a frequency corresponding to the unbinding rate of the crosslinking protein alpha-actinin (Fig. 2C) [107]. A maximum in $G''(f)$ and corresponding turnover in $G'(f)$ is predicted to occur when the network components relax significantly quicker than the unbinding of crosslinking elements [107], implying that this approach could be applied in OT-MR studies of semiflexible protein networks coupled via transient crosslinks to determine unbinding kinetics.

**Future prospects**

Experimentally, much has been learned about protein mechanics and assembly by using optical tweezers in both SMFS and MR configurations. Alternative arrangements of optical traps have the potential to provide distinct insight into the study of protein interactions and assembly. For example, line optical tweezers could be used to study kinetics of transient protein-protein interactions important for nucleation of higher-order assembly, by studying colloidal binding kinetics arising from protein bridges [108]. The use of holographic optical tweezers capable of positioning and quantifying the response of large numbers of particles could be used to determine how network assembly or remodelling is coupled through space and time. Simultaneous measurements at multiple locations throughout the sample during assembly would also provide mechanical insight into how higher-order structure formation percolates throughout the sample [91, 109]. Development of new particles for manipulation can pave the way to higher-force fibre-stretching measurements without the need for higher laser intensities [110], mitigating concerns about local sample heating. The recent demonstration that nanoscale particles can be used for picoNewton force measurements in OT [111] provides many potential advantages to the study of protein mechanics including faster response times, smaller probe surface area, and access within protein networks that have smaller pore sizes.

Integrating other measurement modalities into optical tweezers are likely to provide a deeper mechanistic understanding of protein mechanics and assembly. The incorporation of confocal fluorescence microscopy into bulk rheometers has enabled studies of network formation, reorganization, and fracture in response to macroscopic strains [50, 109, 112]; similar fluorescence imaging approaches integrated into OT-MR would allow mapping of network changes in response to local perturbations, similar to what has been done using phase-contrast imaging during network assembly [95]. SMFS experiments that incorporate OT and single-molecule fluorescence detection are providing insight into protein-protein [35] and DNA-protein [113, 114] interactions, as well as structural force-induced changes of biomolecular structure [17, 18]. Care must be taken when integrating single-molecule fluorescence detection into OT instruments, both to achieve high levels of fluorescence detection sensitivity, and, for short protein substrates, to avoid the desired fluorescence signal being overwhelmed by autofluorescence from the particles used for manipulation [35]. Integrating microfluidics into OT experiments on protein mechanics allows rapid changes of solution environment, enabling studies of chemically triggered network assembly and disassembly [91].

Quantifying and rationalizing the mechanics of proteins at various hierarchical scales is critical in fields including biomaterials design [1, 115], neurodegeneration [43], active matter [116-119], cellular biology – including the rapidly developing field of liquid-liquid phase separation [120-124], and mechanobiology [125, 126]. Optical tweezers are well suited to probing protein mechanics at scales ranging from single molecules to fibres to networks, and, with the integration of complementary measurement modalities, will continue to deliver new insight into the mechanisms by which mechanical responsiveness is imparted by proteins.

**Acknowledgements**

The authors thank the following organizations for funding: Deutsche Forschungsgemeinschaft (DFG; postdoctoral fellowship to KL; Project ID 415037474), the Fonds de recherche du Québec – Nature et technologies (FRQNT; postdoctoral fellowship to MS), and the Natural Sciences and Engineering Research Council of Canada (NSERC; Discovery Grant to NRF).

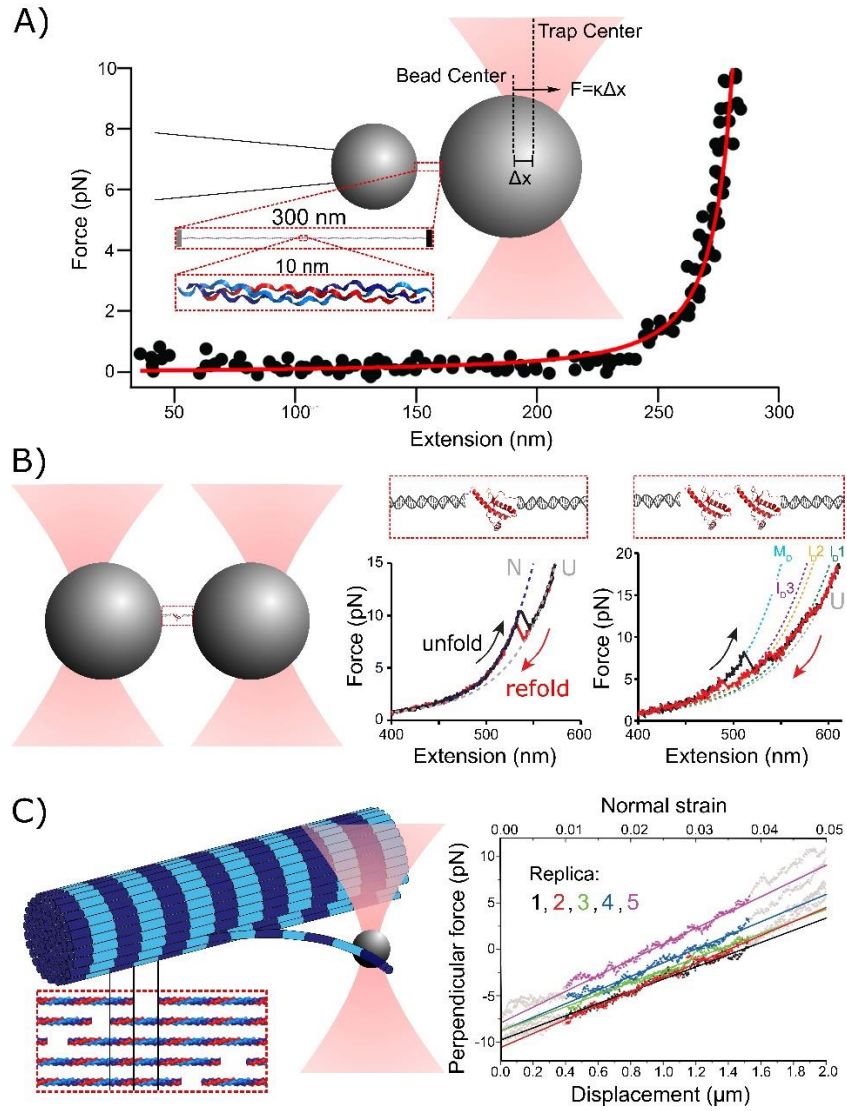

**Fig. 1:** Principles of single-molecule force spectroscopy (SMFS) with optical tweezers (OT). **A)** Schematic of an optical tweezers stretching experiment with collagen. The displacement of the bead center from the trap center Δ$x$ and the trap stiffness $κ$ provide the force applied to stretch the molecule. A Worm-Like Chain (WLC) model (red curve) can be used to fit the resulting force extension curve (black dots). Adapted from [24] with permission. **B)** Experimental scheme for OT-SMFS experiments with short proteins. Prion proteins (PrP) are tethered to polystyrene beads via DNA handles (left). The PrP unfolds and refolds to its native state, dependent on the applied force, as a two-state system (middle). PrP dimers linked at their termini lead to complex force extension curves with multiple intermediates, and more remarkably, adopt a misfolded dimer structure at low force rather than two independently folded domains (right). Adapted from Figure 2 in [41] with permission. **C)** Illustration of measurements of the bending modulus of a collagen fibril. The inset illustrates the highly ordered lateral organization of collagen molecules within a single fibril, which creates a characteristic "D-banding" pattern (dark/light stripes). The optically trapped bead is used to apply bending deformations to a fibril (left). The resulting force-displacement curve reveals the force required for different applied lateral bending strains (right). Adapted from [63] with permission under CC BY licence.

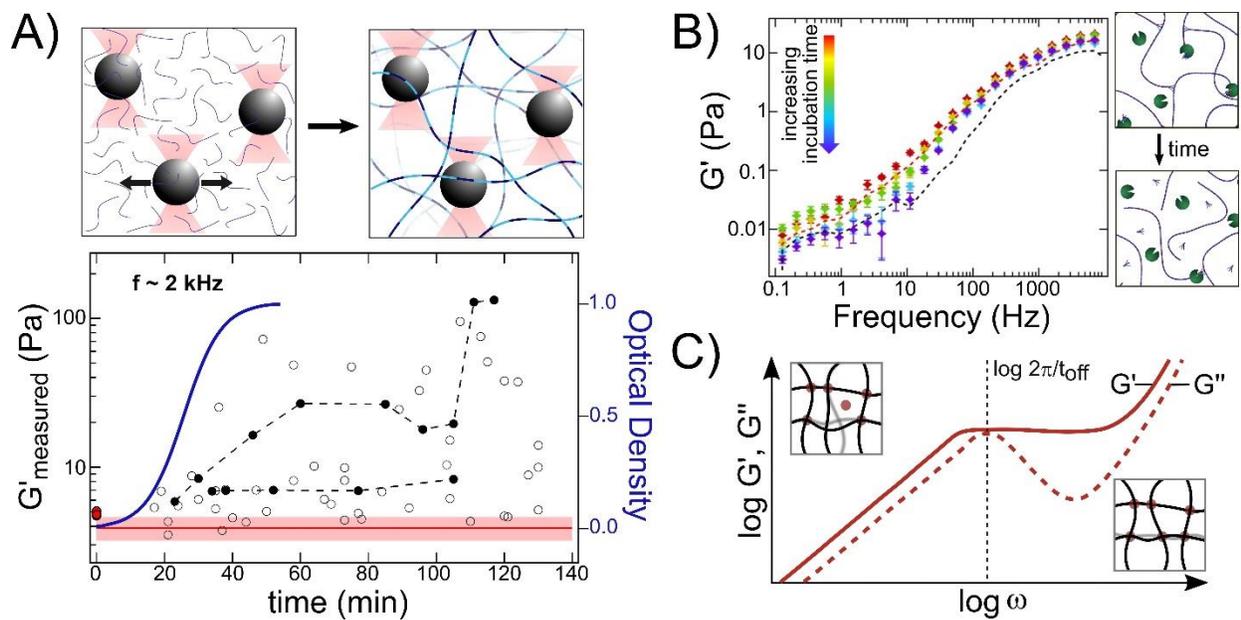

**Fig. 2:** Optical-tweezers based microrheology (OT-MR) measurements of protein network formation and remodelling. **A)** Analysis of trapped bead motion is used to determine how the microscale viscoelastic environment changes during assembly and growth of collagen fibrillar networks (schematic upper panel). Arrows indicate thermally driven fluctuations of the particles within the optical traps, used in passive MR experiments to determine the complex shear modulus of the surroundings. For collagen, it was found that elastic moduli and their spread tend to increase during assembly (lower panel, circles), as the local environment becomes more heterogeneous. The red line and shaded region indicate the range of optical trap elastic moduli $G'_{trap}$ measured for this bead size, while the red dots at zero time indicate the elastic moduli of trap + collagen solutions in acidic conditions, where assembly cannot occur. Filled black circles indicate repeated measurements on the same bead at multiple times during assembly, illustrating distinct evolutions of local mechanics. These experiments found $G'_{max}$ (at fixed frequencies) to increase with the same sigmoidal kinetics as the optical turbidity used to measure growth of the network (blue curve). Adapted from [79] with permission under [CC BY licence](CC BY licence). **B)** OT-MR probes the effect of transient protein-protein interactions that catalyse protein network assembly. Collagen assembly is accelerated by telopeptides, short regions flanking the triple helix (shown as small forked ends in the schematics). Even in acidic conditions where assembly cannot occur, solutions of collagens with intact telopeptides (red dashed line) exhibit a significantly greater $G'$ at low frequencies than collagens with telopeptides enzymatically removed (purple dashed line). The decrease in $G'$ can be detected in real time, as enzymes gradually remove telopeptides (colored markers), thereby reducing protein-protein interactions. Adapted from [106] with permission from Elsevier. **C)** It may also be possible to extract the kinetics of transient crosslinking proteins with OT-MR, as found in bulk rheology experiments on actin. At shorter times / higher frequencies (right inset), only short-range bending fluctuations of the actin filaments can occur, while at longer times /lower frequencies (left inset) actin filaments can undergo larger-scale deformation enabled by unbinding of a crosslinking protein (red circle). Thus, $G''$ of a crosslinked actin gel exhibits a local maximum at a frequency corresponding to the unbinding rate (inversely proportional to the characteristic unbinding time $t_{off}$) of the protein crosslinker. Adapted from [107] with permission (copyright 2010 by the American Physical Society).